\documentclass[floatfix,twocolumn]{revtex4}
\usepackage{graphicx}
\usepackage{dcolumn}
\usepackage{bm}
\usepackage{amssymb}
\usepackage{times}
\usepackage{amsmath}
\usepackage{amsfonts}
\usepackage{mathrsfs,amssymb}
\usepackage{setspace}
\usepackage{epsfig}
\usepackage{latexsym}
\usepackage{bbm}
\usepackage{subfigure}
\usepackage{float}
\usepackage{flafter}
\usepackage{bm}

\begin{document}

\title{Strong correlation induced charge localization in antiferromagnets}
\author{Zheng Zhu$^{1}$, Hong-Chen Jiang$^{2},$ Yang Qi$^{1}$, Chu-Shun Tian$
^{1}$, and Zheng-Yu Weng$^{1\ast }$}
\affiliation{$^{1}$Institute for Advanced Study, Tsinghua University, Beijing, 100084, {\
China}\\
$^2$Kavli Institute for Theoretical Physics, University of California, Santa
Barbara, CA, 93106-4030, U.S.A.\\
$*$ corresponding author, email: weng@tsinghua.edu.cn}

\begin{abstract}
 The fate of an injected hole in a Mott antiferromagnet is an outstanding issue of strongly correlated physics. It provides important insights into doped
Mott insulators closely related to high-temperature superconductivity in
cuprates \cite{Anderson87,Anderson,Lee06}. Here, we report a
systematic numerical study based on the density matrix renormalization
group (DMRG). It reveals a remarkable novelty and surprise for the
single hole's motion in otherwise well-understood Mott insulators. Specifically, we find that the charge of the hole is
self-localized by a novel quantum interference mechanism purely of strong correlation origin \cite{Weng1996,Weng1997}, in contrast
to Anderson localization due to disorders \cite{Anderson58}. The common belief of quasiparticle picture is invalidated by the charge localization concomitant with spin-charge separation: the spin of the doped hole is found to remain a mobile object. Our findings
unveil a new paradigm for doped Mott insulators that emerges already
in the simplest single hole case.
\end{abstract}

\maketitle

\date{\today}


A critical step in understanding the physics of doped Mott insulators is to
investigate the motion of a single hole in Mott antiferromagnets \cite
{Anderson,Lee06,Anderson90PRL,Shraiman88,SCBA1,SCBA2,SCBA3,SCBA4,Dagotto1994,ED,Weng1996,Weng1997,Shih97,White2, Laughlin97,Weng2001,nagaosa04}.
It is easy to imagine that an injected hole in an antiferromagnetically ordered state leaves on its path the trace of spin mismatches in the staggered
magnetization direction (to be called \textquotedblleft $S^{z}$-string\textquotedblright\ below). Such $S^{z}$-string was expected \cite{khomskii} to cause the localization of the hole, but, as was later
realized \cite{SCBA2}, it can be self-healed via quantum spin flips. It was then conjectured \cite{SCBA1,SCBA2,SCBA3,SCBA4}
that a quasiparticle picture should be still valid in long-distance,  as if the Bloch theorem holds true even for such a many-body system. The quasiparticle picture was further supported
by finite-size exact diagonalizations \cite{Dagotto1994} on lattices up to $32$ sites \cite{ED}.  However, experimentally, in angle-resolved photoemission spectroscopy
(ARPES) studies \cite{Shen95,ARPES,Shen04}, very broad single-particle spectral features have
been observed in parent Mott insulator materials such as $\mathrm{Ca}_{2}%
\mathrm{CuO}_{2}\mathrm{Cl}_{2}$ and $\mathrm{Sr}_{2}\mathrm{CuO}_{2}\mathrm{%
Cl}_{2}$. They cast a serious doubt \cite{Laughlin97,Weng2001} on the rationale of understanding an individual hole created by the photon within a
quasiparticle description. Furthermore, the charge localization is a general phenomenon for a lightly doped cuprate before the system becomes a high-$T_{c}$
superconductor. Whether the localization here is intrinsic or due to the presence of disorders remains unsettled.

\begin{figure}[htbp]
\begin{center}
\includegraphics[width=0.45\textwidth]{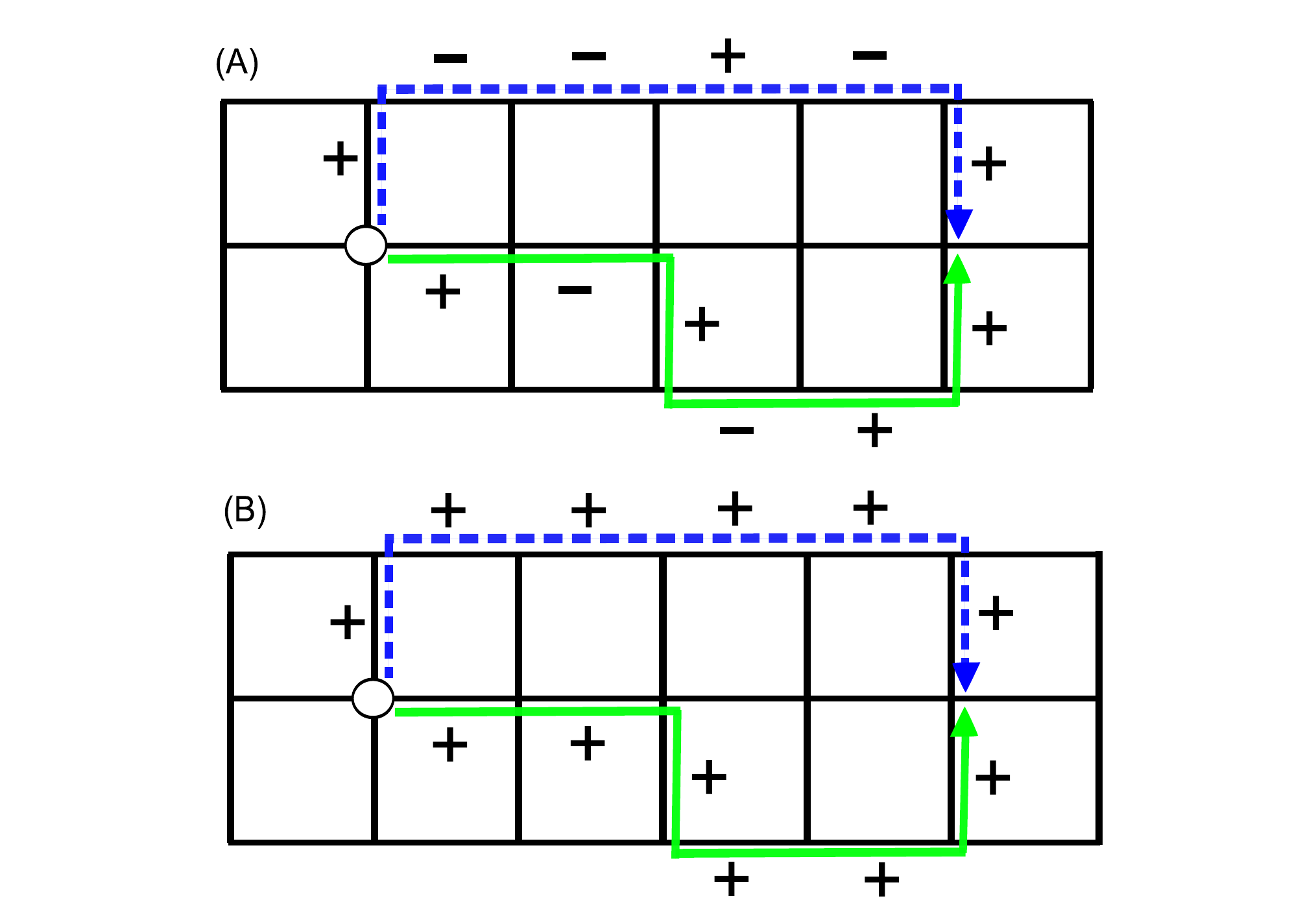}
\end{center}
\par
\renewcommand{\figurename}{Fig.}
\caption{Quatum interference picture of one hole in the $t$-$J$ model (A):
As the hole moves from the injection point to a distant site, a sign
sequence -- the phase string \cite{Weng1996,Weng1997}  -- is left behind as exemplified by the
solid (green) and dashed (blue) lines. Here each {\ $+$ ($-$) sign} faithfully
records an event that the hole exchanges its position with the nearest
neighboring $\uparrow $ ($\downarrow $) spin during its motion. (B) By
contrast, the phase string can be wholly switched off in the $\protect
\sigma\cdot t$-$J$ model (see the text) such that the amplitude associated with each
path is positive definite (of the same absolute magnitude as in the $t$-$J$
model). The presence (absence) of phase strings will drastically alter the
long-wavelength behavior of the hole via quantum destructive (constructive)
interference as demonstrated in this work.}
\label{phase-int}
\end{figure}

Theoretically, the validity of a quasiparticle description has been questioned in the literature as well. It was first argued
by Anderson \cite{Anderson90PRL} on general grounds that the quasiparticle spectral
weight should vanish due to an unrenomalizable many-body phase shift induced
when a hole is injected into a Mott insulator. Later, the discovery of the so-called
phase string \cite{Weng1996,Weng1997} leads to a microscopic justification of this crucial
observation. Here, a phase string may be regarded as an
$S^{\pm}$-string as opposed to the aforementioned
\textquotedblleft $S^{z}$-string\textquotedblright. It is irreparable by quantum spin flips in contrast to the $S^{z}$-string. It
was further predicted \cite{Weng2001} that the phase string, instead of the $S^{z}$-string \cite{khomskii}, is responsible 
for an intrinsic self-localization of the injected hole in two dimensions.
 
In this work, a large-size numerical simulation is used as a powerful machinery to resolve this
issue. This approach is of particular importance because the hole's motion is highly sensitive to
quantum interference developing at long distance. To
this end, we study the ladder systems by the DMRG method \cite{White1992}, in which the length scale
along one direction can be sufficiently large. As a ladder
sample is long enough, we find that the charge is generally localized so long as the leg number is larger than
one. The localization scale decreases monotonically as the leg number
increases, suggesting a stronger self-localization in the two dimensional limit.
Contrary to this, if the sample is not long enough, the injected hole behaves itinerantly, implying that the
quasiparticle picture found in
earlier numerical studies \cite{Dagotto1994,ED} is likely a small-size
effect. We further show that complementary to the charge localization, the
system exhibits novel spin-charge separation. That is, the \textquotedblleft
spinon\textquotedblright\ that carries the spin quantum of the doped hole
remains mobile. We demonstrate that both charge self-localization and spin-charge
separation find their origin in the phase string, a unique property of the $t
$-$J$ model \cite{Weng1996,Weng1997,Wu-Weng-Zaanen}. As a further evidence, we find that all of these exotic properties
disappear once the phase string is turned off artificially in the simulation, and
a well-defined quasiparticle description is recovered in consistency with the common
wisdom \cite{SCBA2} of spin polaron picture.

\textit{Models and methods.---} A prototypical doped Mott insulator is
described by the $t$-$J$ Hamiltonian, $H_{t\text{-}J} = H_t+H_J$, with
\begin{equation}
\begin{split}
H_t &= -t \sum_{\langle {ij}\rangle \sigma } {({c_{i\sigma }^{\dag
}c_{j\sigma }+h.c.})}, \\
H_J &= J \sum_{\langle {ij}\rangle } {(\mathbf{S}_{i}\cdot \mathbf{S}_{j}-%
\frac{1}{4}n_{i}n_{j})}.
\end{split}
\label{a}
\end{equation}%
Here, ${{c_{i\sigma }^{\dagger }}}$ is the electron creation operator at
site $i$, ${\mathbf{S}_{i}}$ the spin operator, and ${n_{i}}$ the number
operator. The summation is over all the nearest-neighbors, $\langle
ij\rangle $. The Hilbert space is constrained by the no-double-occupancy
condition, i.e., $n_{i}\leq 1$. At half-filling, $n_{i}=1$, the system
reduces to Mott insulators (antiferromagnets) with a superexchange coupling,
$J$. Upon doping a hole into this system, $\sum_{i}n_{i}=N-1$ ($N$ the
number of the lattice sites), and the hopping process is triggered as
described by the hopping term, $H_t$, with $t$ the hopping integral.

For the single hole doped $t$-$J$ model in a bipartite lattice, there exists
an exact theorem, which states \cite{Weng1996,Weng1997,Wu-Weng-Zaanen} that
the propagation of the hole is a superposition of quantum amplitudes of
all the paths, with each path carrying a unique sign sequence known as the 
phase string, i.e.,
\begin{equation}  \label{eq:3}
(+1)\times (-1)\times (-1)\times \cdots.
\end{equation}
Here, the sign $\pm $ on the right-hand side keeps track of an $\uparrow$ or
$\downarrow$-spin exchanged with the hole at each step of hopping as
illustrated in {\ Fig.}~\ref{phase-int} (A).

\begin{figure*}[btp]
\includegraphics[width=0.8\textwidth]{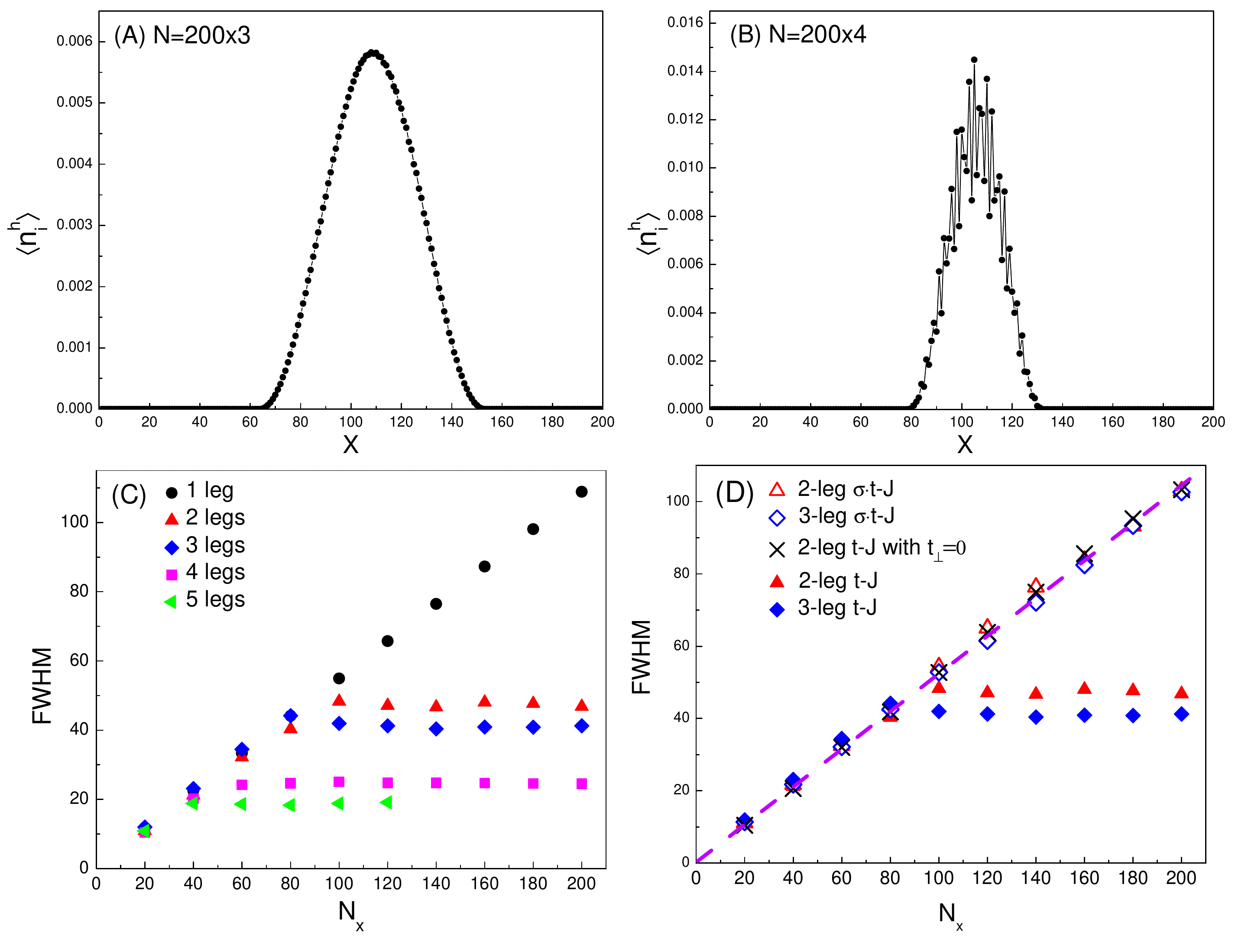} %
\renewcommand{\figurename}{Fig.}
\caption{Self-localization of the charge in a $3$-leg ladder system
of size $N=200 \times 3$ (A) and $4$-leg ladder of $N=200 \times 4$ (B).
The density distribution $\left \langle {n_{i}^{h}}\right \rangle $ along the $x$ axis in a
middle leg is well fitted by the Gaussian function with a full width at half-maximum
(FWHM). (C) The behavior of FWHM depends on
the leg number $N_y$: for the ladders of $N_y > 1$, the FWHM first
increases linearly at small $N_x$ and then saturates at large $N_x$, with a length scale
monotonically decreasing with the leg number; for $N_y = 1$, the FWHM grows
unboundedly as the sample length increases. (D) Delocalization of the charge
in the absence of phase string interference for the two
modified models, namely the $\sigma\cdot t$-$J$ model and
the $t$-$J$ model with interchain hopping coefficient $t_\perp=0$.
The corresponding FWHM unboundedly increases with $N_x$ (open triangles, open diamonds, and crosses).
The dashed line ($=N_x/2$) is a guide to the eye.  }
\label{eg}
\end{figure*}

The quantum interference of phase strings from different paths is
expected to be strong and play the fundamental role in dictating the
long-distance behavior of the hole. In order to uniquely pin down the effect
of phase strings, we also introduce the so-called $\sigma\cdot t$-$J$ model with the
hopping term $H_t$ in Eq. (\ref{a}) replaced by
\begin{equation}
H_{\sigma\cdot t} = -t \sum_{\langle {ij}\rangle \sigma }\sigma {({c_{i\sigma
}^{\dag }c_{j\sigma }+h.c.})},  \label{c}
\end{equation}
(i.e., an extra sign $\sigma=\pm 1$ is added). It is easy to show, following Refs.
\cite{Weng1996,Weng1997,Wu-Weng-Zaanen}, that the phase string (\ref{eq:3})
disappears, whereas the positive weight for each path remains unchanged [as illustrated
in {\ Fig.}~\ref{phase-int} (B)]. In other words, one can artificially
switch on and off the phase string effect between the $t$-$J$ and $\sigma\cdot t$-%
$J$ models to study its novel consequences.

Previously the DMRG algorithm has already been used to study the $t$-$J$
model in ladder systems at low hole doping \cite{White1,White2}. Below
we shall focus on the one hole case on bipartite lattices of $N=N_{x}\times
N_{y}$, where $N_{x}$ and $N_{y}$ are the site numbers in the $x$ and $y$
directions, respectively. By using the DMRG method, we shall study the
ladders with small $N_{y}$ (from $1$ to $5$) and sufficiently large $N_{x}$.
Here we set $J$ as the unit of energy and focus on the $t/J=3$ case unless
otherwise specifically stated. Enough numbers of states in each DMRG block
are kept to achieve a good convergence with total
truncation error of the order of $10^{-8}-10^{-12}$.

\emph{Self-localization of the charge.---} Typical examples of the hole density
distribution, $\left \langle {n_{i}^{h}}\right \rangle \equiv
1-\left
\langle {n_{i}}\right \rangle $, are shown in {\ Fig.}~\ref{eg} (A)
and (B) for $N_{y}=3$ and $4$, respectively. Here the charge profiles are plotted along the $x$ direction for a middle leg of the ladders. They are localized in the
central region of the ladders with open boundaries. Upon summing up the distribution at all the sites of different legs, the sum rule: $\sum_{i}\left \langle {n_{i}^{h}}\right
\rangle =1$ is satisfied. Examples of the contour plot of $\left \langle {
n_{i}^{h}}\right \rangle $ in the $x$-$y$ plane can be found in Fig.~\ref
{Total_Contour} in Supplementary Information.

One may systematically change the ladder length $N_x$ to study the full
width at half-maximum (FWHM) of the charge profile. {\ Fig.}~\ref{eg} (C) shows that for
ladders with different $N_{y}$, the FWHM increases linearly for small sample
lengths, while saturates for sufficiently large sample lengths (except the $%
N_{y}=1$ case). The former indicates that the doped hole still remains itinerant at a smaller $N_x$. The latter suggests that a self-localization takes place for the doped hole, when the ladder length is long enough where the FWHM is no longer sensitive to the
boundaries along the $x$-direction. Later on, by a more precise analysis, it will be shown that such localization only involves the charge of the doped hole, not its spin.
Furthermore, {\ Fig.}~\ref{eg} (C) clearly shows that
the saturated FWHM at $N_{y}>1$ monotonically decreases with the increase of
the leg number $N_{y}$, implying even stronger localization in the
two-dimensional limit. By contrast, there is no indication of saturation in
FWHM for long one-dimensional chains ($N_y=1$, with $N_{x}$ up to $500$),
which is consistent with the fact that the doped hole in strictly one
dimension exhibits the Luttinger liquid behavior \cite{Anderson}.

The destructive quantum interference effect of phase strings in the $t$-$J$ model
provides a natural explanation for the charge \textquotedblleft self-localization\textquotedblright\ observed in
the above DMRG results. To further confirm the phase string origin of
self-localization, we switch off the phase string interference by two
methods and then repeat the previous procedure of numerical simulations. In
the first method, we study the $\sigma\cdot t$-$J$ model introduced above. In
the second method, we set the interchain hopping coefficient, $t_\perp$, to be
zero such that the hole can only move in the $x$-direction as if in a
one-dimensional chain. In both cases, the interference picture shown in {\
Fig.~\ref{phase-int} (A) } breaks down. The numerical results for these two
models are presented in {\ Fig.}~\ref{eg} (D) for two- and three-leg cases,
in which the FWHM of the hole density distribution restores the behavior of linear
increase with the sample length. It indicates that the charge becomes
delocalized in both cases, in sharp contrast to the self-localization in
the $t$-$J$ model.

Although the observed self-localization is insensitive to the parity
(even-odd) of the leg number, the hole {\ distribution is}, as shown in
Fig.~\ref{eg} (A) and (B) and Supplementary Information: for the even-leg
ladders ($N_{y}=2,\,4$), there are always small {\ spatial} oscillations on
top of the Gaussian density profile, while they are absent for the odd-leg
ladders ($N_{y}=3$, $5$). This can be attributed to distinct decaying
behavior of spin-spin correlations for the odd- and even-leg ladders at
half-filling: the former (latter) follows a power (an exponential) law
reflecting the absence (presence) of spin gap, see Supplementary Information
for further explanations. Furthermore, the distinction of the even-odd
effect disappears with the increase of $t/J$ ratio (see {\ Fig.}~\ref%
{ratio-hole}), indicating that the detailed spin dynamics, governed by $J$,
is not essential to the charge self-localization effect.

\begin{figure}[tbp]
\begin{center}
\includegraphics[width=0.45\textwidth]{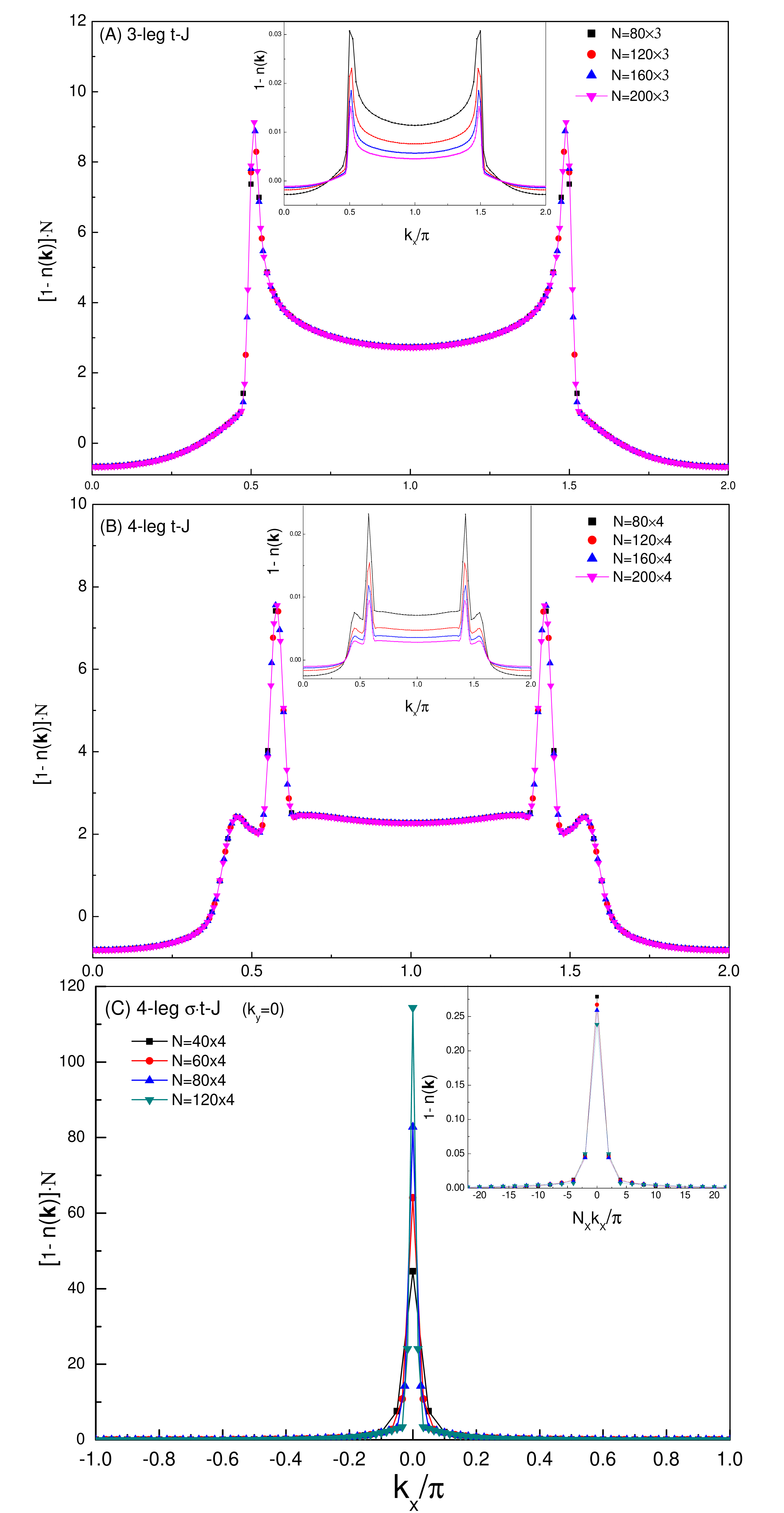}
\end{center}
\par
\renewcommand{\figurename}{Fig.}
\caption{Momentum distribution. For both $3$-leg (A) and $4$-leg (B)
ladders, the momentum distribution of the hole exhibits scaling behavior:
after the rescaling, $1-n(\mathbf{k}) \rightarrow [1-n(\mathbf{k})]N$, the
curves at different sample lengths in the inset collapse onto a single one
represented in the corresponding main panel. Note that {\ we fix $k_y=\frac{%
2\protect \pi}{3}$ for the $3$-leg case and $k_y=\frac{\protect \pi}{2}$ for
the $4$-leg case}, and plot the distribution along $k_x$. Representive
contour plots in the whole $k_x$-$k_y$ plane are shown in {\ Fig.}~\protect
\ref{n(k)contour} (Supplementary Information). (C) The quasiparticle picture is recovered
for the $\sigma\cdot t$-$J$ model. The main panel shows that the momentum distributions of
different sample lengths collapse to a single momentum at $k_x=k_y=0$ for the 4-leg ladder; The inset shows that a universal
curve is obtained for $1-n({\bf k})$ after the rescaling:
$k_x \rightarrow N_x k_x$, which suggests a nonvanishing quasiparticle spectral weight peaked at ${\bf k}=0$. }
\label{collapse}
\end{figure}

\emph{Momentum distribution.---} Now let us examine the momentum
distribution of the hole by studying $n(\mathbf{k})\equiv \sum_{\sigma }\langle {c_{\mathbf{%
k}\sigma }^{\dag }c_{\mathbf{k}\sigma }}\rangle$, which can be obtained by a
Fourier transformation of $\sum_{\sigma }\langle {c_{i\sigma }^{\dag
}c_{j\sigma }}\rangle$. At half-filling, one finds $n(\mathbf{k})=1$ and for
the one hole case, $1-n(\mathbf{k})$ is shown in {\ Fig.}~\ref{collapse}.

The insets of {\ Fig.}~\ref{collapse} (A) and (B) present the hole momentum
distribution $1-n(\mathbf{k})$ as a function of $k_{x}$ for fixed $%
k_{y}=2\pi /3$ in the $3$-leg ladder and $k_{y}=\pi /2$ in the $4$-leg
ladder, respectively. The value of $k_{y}$ is chosen in a way that the \textquotedblleft sudden change\textquotedblright\
in $1-n(\mathbf{k})$ can reach maxima by varying $k_{x}$,
according to the contour plots in the $k_{x}$-$k_{y}$ plane (see
Supplementary Information). A very interesting feature is that after the rescaling: $%
1-n(\mathbf{k}) \rightarrow [ 1-n(\mathbf{k})] N$, all the curves in the
inset of {\ Fig.}~\ref{collapse}(A) [or (B)] collapse onto a universal curve
shown in the corresponding main panel. If one defines the Fermi surface by
the sudden jump in the momentum distribution function, then the two
universal curves suggest that the quasiparticle spectral weight is upper-bounded by $1/N$
for large $N$, and vanishes in the thermodynamic limit.

To satisfy the sum rule, $\sum_{\mathbf{k}}(1-n(\mathbf{k}))=1$, the width
of the jump in $1-n(\mathbf{k})$ must remain finite in the limit $%
N\rightarrow \infty $. This is clearly shown in {\ Fig.} \ref{collapse} (A)
and (B), consistent with a finite localization length in the real space.
We have also calculated the hole
momentum distribution at different ratios of $t/J$. As shown in
Supplementary Information, for a given sample size, the jump near the
Fermi point is continuously reduced with increasing $t/J$, which is
qualitatively consistent with earlier work \cite{Dagotto1994}.

Again a sharp contrast arises once the phase string is turned off in the $%
\sigma\cdot t$-$J$ model, as shown in Fig. \ref{collapse}(C) for the 4-leg ladder. Here the wide
spread of $1-n(\mathbf{k})$ in the momentum space seen in (A) and (B)
collapses into a sharp peak at $k_x, k_y=0$. In particular, a universal
curve (the inset) is obtained by the rescaling: $k_x \rightarrow N_x k_x$ {\
instead of} by $1-n(\mathbf{\mathit{k}}) \rightarrow [ 1-n(\mathbf{k})] N$
as done in the main panel. This clearly indicates that the quasiparticle
spectral weight is finite and a Bloch-like state is restored with a definite
momentum at $\mathbf{k}=0$ (a peak at the equivalent of $k_x, k_y=\pi$ is
also found).

\begin{figure*}[tbp]
\begin{center}
\includegraphics[width=0.8\textwidth]{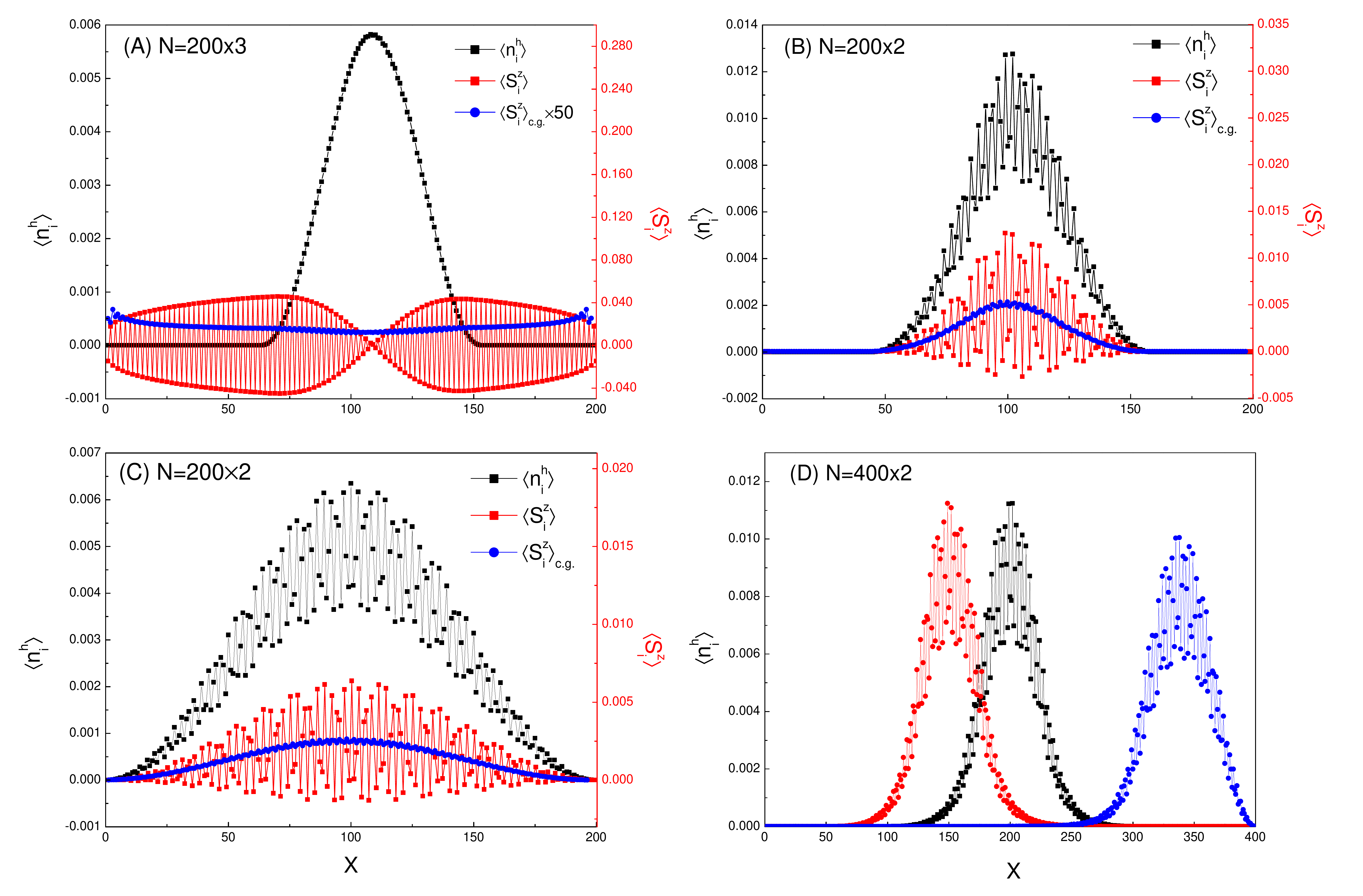}
\end{center}
\par
\renewcommand{\figurename}{Fig.}
\caption{(A) Spin-charge separation in a 3-leg ladder with a $\downarrow $ spin electron removed. The charge of the hole is localized at the sample center and the spin of $S^{z}=1/2$ spreads over the entire sample, with the coarse grained density $\langle S_{i}^{z}\rangle _{c.g.}$ approximately uniform. (B) In a 2-leg ladder, the charge and spin of the doped hole remain in the same localized region because of a finite spin-gap in the spin background. (C) An extended profile of the charge can be obtained with further reducing DMRG truncation error (see the text); In this case, although the charge remains localized, the mobile spinon gains a tiny additional energy (cf. Fig.~\ref{E_dif}). (D) In a $N=400\times 2$ ladder, by adding weak chemical potentials the hole density can be localized anywhere along the ladder.}
\label{scseparation}
\end{figure*}

\emph{Spin-charge separation.---} Complementary to
the charge localization illustrated in {Fig.}~\ref{eg}, we further explore
the fate of the spin-$1/2$ associated with the hole injected into the
half-filled spin-singlet background.

For the $3$-leg ladder, the existence of spin-charge separation is explicitly
demonstrated by {Fig.}~\ref{scseparation}(A). Here the single
hole is created by removing a $\downarrow $ spin electron.  While the charge
of the doped hole is well localized at the sample center, the extra
spin of $S^z=1/2$ spreads over the sample. It is indicated by spin density $\left \langle S_{i}^{z}\right \rangle $ as well as the coarse
grained $\left \langle S_{i}^{z}\right \rangle _{c.g.}$, which is  
approximately uniform across the ladder.  This is probably the most
direct display of electron fractionalization for a doped Mott insulator,
thanks to the charge localization. Note that it is
very different from the well-known one-dimensional Luttinger liquid, where both charge  (holon) and spin (spinon) are mobile \cite {Anderson}. 

For an even-leg ladder, the presence of a spin gap in the spin background
renders the spin-charge separation less explicit because the
extra spin of $S^{z}=1/2$ will also tend to stay in the hole region in order
to reduce its superexchange energy cost, as illustrated in {Fig.}~\ref{scseparation}(B) 
for the $2$-leg ladder case. To directly observe a
similar spin-charge separation picture in an even-leg ladder, the spin gap
has to be small enough, a condition that can be achieved
only if the leg number is sufficiently large. In this limit, the behavior of the single
hole in both even- and odd-leg ladders is expected to converge with a stronger charge localization.

Nevertheless, the spin-charge separation in the $2$-leg ladder can be well confirmed by a different
method. Note that, first, if the charge part -- namely the holon -- is
localized, the localization center can be anywhere in a translationally invariant
system. In particular, an \textquotedblleft extended\textquotedblright \
profile as a linear superposition of the localized wave packets at different
positions should be also energetically degenerate. Second, if {Fig.}~\ref%
{scseparation}(B) really corresponds to spin-charge separation, the extra spin-$1/2$ --
namely the spinon -- should be more or less \textquotedblleft
free\textquotedblright \ within the regime
expanded by the charge distribution. Then the spinon will prefer a more
extended distribution of the hole to further lower its
energy. In other words, the degeneracy of localized states should be lifted in
favor of an extended profile in order to gain additional spinon energy. 

\begin{figure}[tbp]
\begin{center}
\includegraphics[width=0.53\textwidth]{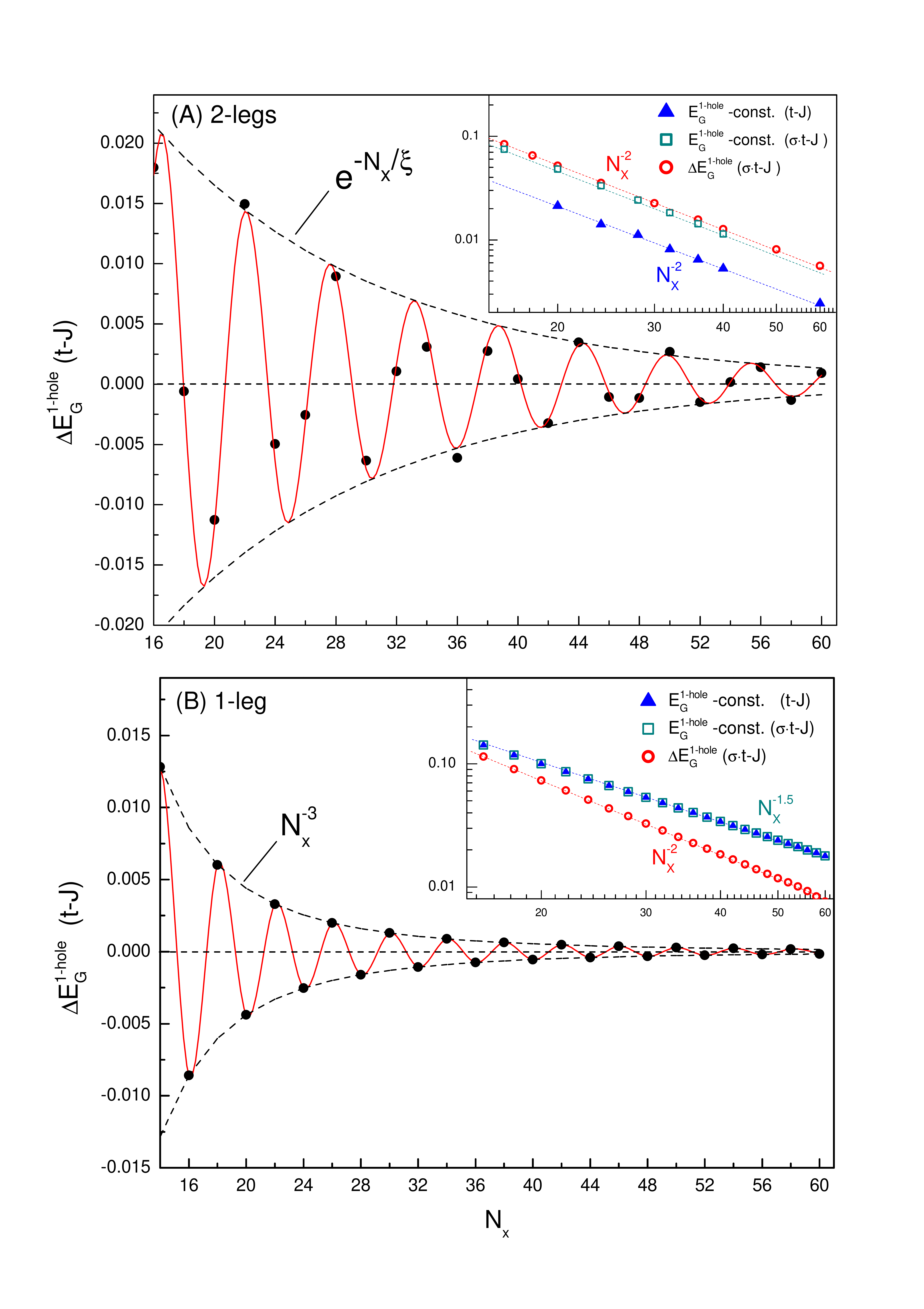}
\end{center}
\par
\renewcommand{\figurename}{Fig.}
\caption{The energy difference $\Delta E_{G}^{\text{1-hole}}$ between anti-periodic and periodic boundary conditions along the x-direction for the charge shows an oscillation vs. $N_x$, which is a direct manifestation of the phase string effect.  The exponential decay of the envelop (dashed curves) of $\Delta E_{G}^{\text{1-hole}}$ shows the charge localization for the 2-leg ladder (A), while the power law decay for the 1-leg ladder case (B) indicates a Luttinger liquid behavior. By sharp contrast,  in the inset of (A), a completely different finite-size behavior ($1 /N_{x}^{2}$ ) is exhibited in the ground state energy $E_{G}^{\text{1-hole}}$ for the 2-leg ladder, confirming the spin-charge separation. For comparison, both $\Delta E_{G}^{\text{1-hole}}$ and $E_{G}^{\text{1-hole}}$ follow the same $1/N_{x}^{2}$ dependence  once the phase string is turned off, as shown in the inset of (A), where the quasiparticle picture is restored with a spin-charge recombination.  In the inset of (B), similar comparisons for the 1-leg case are presented.   }
\label{E_dif}
\end{figure}

Indeed, by increasing DMRG sweep number with the truncation error
reaching much lower than $10^{-8}$, the localized profile in {Fig.}~\ref%
{scseparation}(B) is found to slowly converge to an extended one as shown in
(C) with the total energy lowered by $\sim 10^{-4}J$. To identify the nature of this small energy gain, one may vary the sample size from $N_{x}=16$ to $800$, and repeat the same procedure to obtain the one-hole energy $E_{G}^{\text{1-hole}}$ at each $N_{x}$ [with the
same full convergence to the extended charge profile as in {Fig.}~\ref%
{scseparation}(C)]. As shown in the inset of {Fig.}~\ref%
{E_dif}(A), with a constant term excluded, $E_{G}^{%
\text{1-hole}}$ is well fitted by $\alpha (\pi /N_{x})^{2}$ (blue triangles) with the
coefficient $\alpha =0.87 J$, as if it is contributed by some \textquotedblleft
free particle\textquotedblright\ in a box of length $N_x$ \cite{White}. 

Next, we show that such $1 /N_{x}^{2}$ (finite-size) contribution to $E_{G}^{\text{1-hole}}$ is not from the charge sector. For this purpose, we make the $2$-leg ladder as
a ribbon by connecting the boundaries on the two edges of the ladder in
the $x$-direction. Then we calculate the energy difference $\Delta E_{G}^{%
\text{1-hole}}\equiv E_{G}^{\text{1-hole}}(\pi )-E_{G}^{\text{1-hole}}(0)$
with threading through a flux $\pi $ into the ribbon. It corresponds to the change of 
boundary condition from the periodic to anti-periodic one for the hopping
term and is equivalent to coupling a small electric field to the charge
sector.  As shown in {Fig.}~\ref {E_dif}(A), $\Delta E_{G}^{\text{1-hole}}$
oscillates strongly and falls off exponentially as $e^{-N_x/\xi}$ with $\xi\sim 14.5$. It behaves completely different from $1 /N_{x}^{2}$ behavior exhibited in $E_{G}^{\text{1-hole}}$, indicating that
the charge indeed remains localized. The \textquotedblleft
free particle-like\textquotedblright \ behavior in $E_{G}^{\text{1-hole}}$ should therefore
solely originate from the charge-neutral part of the doped hole, i.e., the spinon. 

By a sharp contrast, upon turning off the phase-string in the $\sigma\cdot t$-$J$ model, both $E_{G}^{\text{1-hole}}$ and $\Delta E_{G}^{\text{1-hole}} $ follow the same $1 /N_{x}^{2}$ behavior as clearly shown in the inset of {Fig.}~\ref {E_dif}(A). It is consistent with the previous conclusion that in this case the doped hole is restored as a quasiparticle, which carries both charge and spin and thus responds very similarly to the boundary condition (i.e., the finite-size effect) as a single mobile object.

The spin-charge separation is well-known for the one-dimensional chain \cite {Anderson}.  {Figure}~\ref%
{E_dif}(B) presents the corresponding $1$-leg results. As shown in the inset, the finite-size effect of $E_{G}^{\text{1-hole}}$ ($\propto 1 /N_{x}^{1.5}$) is indistinguishable from the phase string free case (with open boundary conditions). But the charge response of $\Delta E_{G}^{\text{1-hole}} $ is again totally different [the main panel of {Fig.}~\ref {E_dif}(B)]. It oscillates strongly as a function of $N_x$, similar to the $2$-leg case in {Fig.}~\ref {E_dif}(A). As a matter of fact, it can be directly attributed to the phase string modulation in the expression for the ground state energy given in Ref. \cite{Weng2001}. The main difference from the $2$-leg case is that the envelop of $\Delta E_{G}^{\text{1-hole}}$ decays with increasing $N_x$ in a power law fashion ($\propto 1 /N_{x}^{3}$) instead of an exponential one. This is consistent with the fact that the holon is not localized in the one-dimensional case. It is interesting to notice that even without the phase string effect, $E_{G}^{\text{1-hole}}$ and $\Delta E_{G}^{\text{1-hole}} $ behave distinctly for the $\sigma\cdot  t$-$J$ model. The former $\propto 1 /N_{x}^{1.5}$, while the latter $\propto 1 /N_{x}^{2}$, also indicating a spin-charge separation although the distinction is much less dramatic as compared to the $t$-$J$ model. 

The presence of spin-charge separation enriches substantially the behavior of the hole localized states in the $t$-$J$ ladder systems.  Due to the residual spin-charge interaction, the degenerate charge localized states are superposed to form a more \textquotedblleft
extended \textquotedblright \ distribution to minimize the spinon energy.  It has been shown \cite{HCJ} that generally the DMRG method tends to pick up a \emph{minimally entangled} state among quasi-degenerate states. Thus, spatially localized ones such as those in  {\ Fig.}~\ref{eg} are naturally obtained because a further spinon energy gain for a bigger FWHM would be too tiny to be detected with a given small DMRG truncation error. It also explains why the FWHM of charge distribution shown in {\ Figs.}~\ref{eg} and \ref {scseparation} is usually bigger than the intrinsic localization length $\xi$ as exhibited by $\Delta E_{G}^{\text{1-hole}} $. But the former is still valid to characterize the general trend of the localization, say, as a function of the leg number in  {\ Fig.}~\ref{eg}. On the other hand, by adding some weak local chemical potentials, the quasi-degeneracy of the hole states can be lifted such that the hole density profile is truly localized around the impurities at different locations [as indicated by different colored curves in {Fig.}~\ref {scseparation}(D) for a $N=400\times 2$ ladder]. Here the increase of the total energy of the original $t$-$J$ Hamiltonian is weak (about $10^{-4}J$), presumably from the spinon part as the FWHM is still larger than $\xi$. 

\emph{Discussions.---} Phase string,  charge
localization, and spin-charge separation are
striking results of a single-hole-doped Mott insulator. They are truly incompatible with a quasiparticle description. However, for their novel effects to get fully unveiled, a large sample is needed. Instead of tackling directly a two-dimensional square lattice, in this work, we
have chosen ladders that DMRG can access. As it turns out, very surprising consequences
do show up so long as the ladders are long enough,
\emph{and}, there are more than one paths to realize quantum interference, e.g.,
in a two-leg system.

The most essential physics underlying the observed exotic phenomena is that the motion of the single hole acquires a sequence of signs that precisely and permanently
record each microscopic \emph{hole-hopping-spin-backflow} event in the quantum spin
background. These irreparable phase strings cause singular and destructive interference of quantum waves of the charge. The longer the paths are, the stronger the destructive interference is. The present results
show unequivocally that the self-localization of the charge is a combined effect of quantum interference and multiple scattering between the doped charge and spin background, in
analogy to Anderson localization \cite{Anderson58}.  Contrary to this, the charge remains delocalized in the single chain case due to the lack of quantum interference involving \emph{different} paths. 

The charge localization is further found to be concomitant with spin-charge separation for the two and more-leg ladders. Whereas the phase string renders kinetic energy suppression of the charge, the spin associated with the doped hole remains unfrustrated and thus maintains a coherent motion as a spinon. This has been clearly demonstrated by the distinct finite-size effects of the charge and spin degrees of freedom. In this sense, the electron fractionalization found in the single hole case can be attributed to the energetic reason. Note that the cause for charge localization, i.e., the nontrivial phase string effect, requires an uncertainty of the spin quantum associated with the spin backflow (leading to the $\pm$ signs).  If \emph{each} spin backflow could maintain a definite spin polarization
direction,  the phase string effect would become trivial. Simultaneously the hole would also carry a definite spin as a quasiparticle. But such a spin-charge recombination, being extremely tight in space, would  cost too much kinetic energy. 

This is fundamentally different from the premise for a spin polaron picture -- a
conventional wisdom in which the disturbance of the hole hopping to the spin background is
described by virtual spin excitations that eventually relax back
after the hole moves away. In fact, the spin-polaron or quasiparticle behavior has been
shown to be recovered once the phase string is \textquotedblleft switched
off\textquotedblright \ in the so-called $\sigma\cdot t$-$J$ model, where
everything seems \textquotedblleft normal\textquotedblright \ in a
conventional sense. Here a spin-charge recombination (not too tight spatially) with a good kinetic energy for the quasiparticle can be realized since there is no more singular phase string effect to strongly suppress the kinetic energy of the charge.

Our results have far-reaching consequences. First of all, they suggest that a new state of quantum matter may naturally emerge in sufficiently low doping. Such a state is insulating because of the self-localization of doped charges, not because of a charge gap like in the undoped case. This is a novel non-Anderson-type localization phenomenon since external disorders are absent. Secondly, the strong suppression of the kinetic energy results in a large quasi-degeneracy in the low-energy states. It is expected to lead to a \textquotedblleft glassy\textquotedblright\ behavior in the presence of weak external impurities and a \textquotedblleft competing-order\textquotedblright\ phenomenon in the presence of additional weak interactions. It also provides a possible explanation for the strongly localized state induced by a surface defect observed in the scanning tunneling microscope experiment on $\mathrm{Ca}_{2}%
\mathrm{CuO}_{2}\mathrm{Cl}_{2}$ \cite{STM}. Thirdly, the spin background remains essentially the same as at half-filling,  but the spinons fractionalized from the doped holes will play an important role to determine the low-energy, long-wavelength sector of spin dynamics in the lightly doping. Finally, given that phase strings have been proved \cite{Wu-Weng-Zaanen}
rigorously as the sole sign structure in the $t$-$J$ model on bipartite
lattices, regardless of doping concentration, temperature, and dimensions, 
the present work also provides significant insights into the long-standing
issue of high temperature
superconductivity at finite doping.

\newpage

\begin{widetext}

\section*{Supplementary Information}
\renewcommand{\thefigure}{S\arabic{figure}}
\renewcommand{\theequation}{S\arabic{equation}}
\renewcommand{\thesection}{\Roman{section}}
\setcounter{section}{0}
\setcounter{figure}{0}
\setcounter{equation}{0}

In this supplemental material, we present some technical details.

{\it Spin-spin correlations at half-filling.} ---
At half-filling, the $t$-$J$ model reduces to the Heisenberg model.
For the isotropic Heisenberg coupled-chain systems, the behavior of the even-leg
ladders is dramatically different from that of the odd ones. A well-known fact is that
the even-leg ladders have a spin gap while the odd-leg ladders are
gapless. The former leads to an exponential decay of the spin-spin correlation
while the latter to a power law decay (see
Fig.~\ref{spincorrelation}). The spin gap for the even-leg ladders is
expected to vanish in the large leg number (namely two-dimensional) limit.
Indeed, the spin structure
factor in the $4$-leg case has already exhibited strong antiferromagnetic
correlations as shown in Fig.~\ref{structurefactor}. These
results are consistent with earlier DMRG work\cite{White94}.

\begin{figure}
\begin{center}
\includegraphics[width=0.9\textwidth]{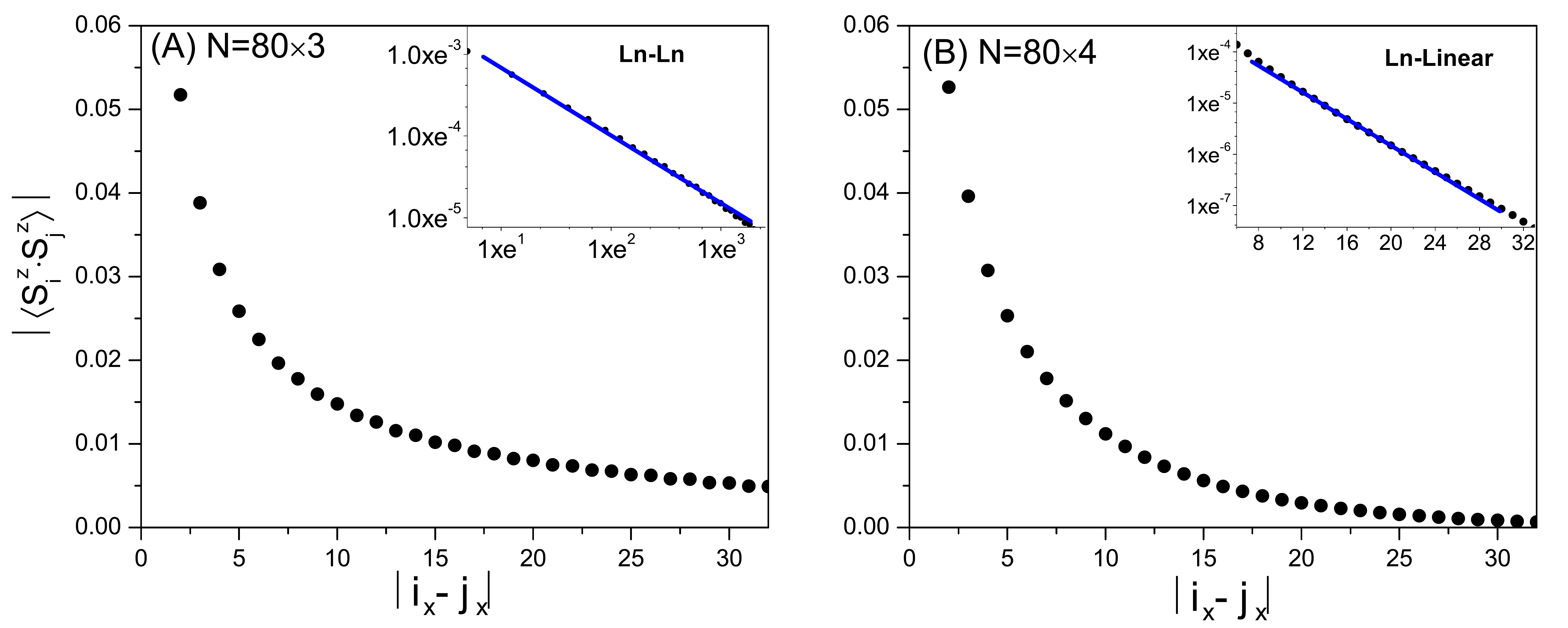}
\end{center}
\par
\renewcommand{\figurename}{Fig.}
\caption{The main panels are the spin-spin correlation $\left|
{\left \langle {S_i^z S_j^z } \right \rangle } \right|$
in the $3$-leg (A) and $4$-leg (B) ladder, respectively.
$i_x$ and $j_x$ are chosen to be on the middle leg of the corresponding ladder. The
insets are $\ln$-$\ln$ plot (A) and semi-$\ln$ plot (B). They are well fitted by
straight (blue) lines, indicating a power law decay
for the $3$-leg ladder and an exponential decay for the $4$-leg ladder.}
\label{spincorrelation}
\end{figure}

\begin{figure}
\begin{center}
\includegraphics[width=10cm,height=6cm]{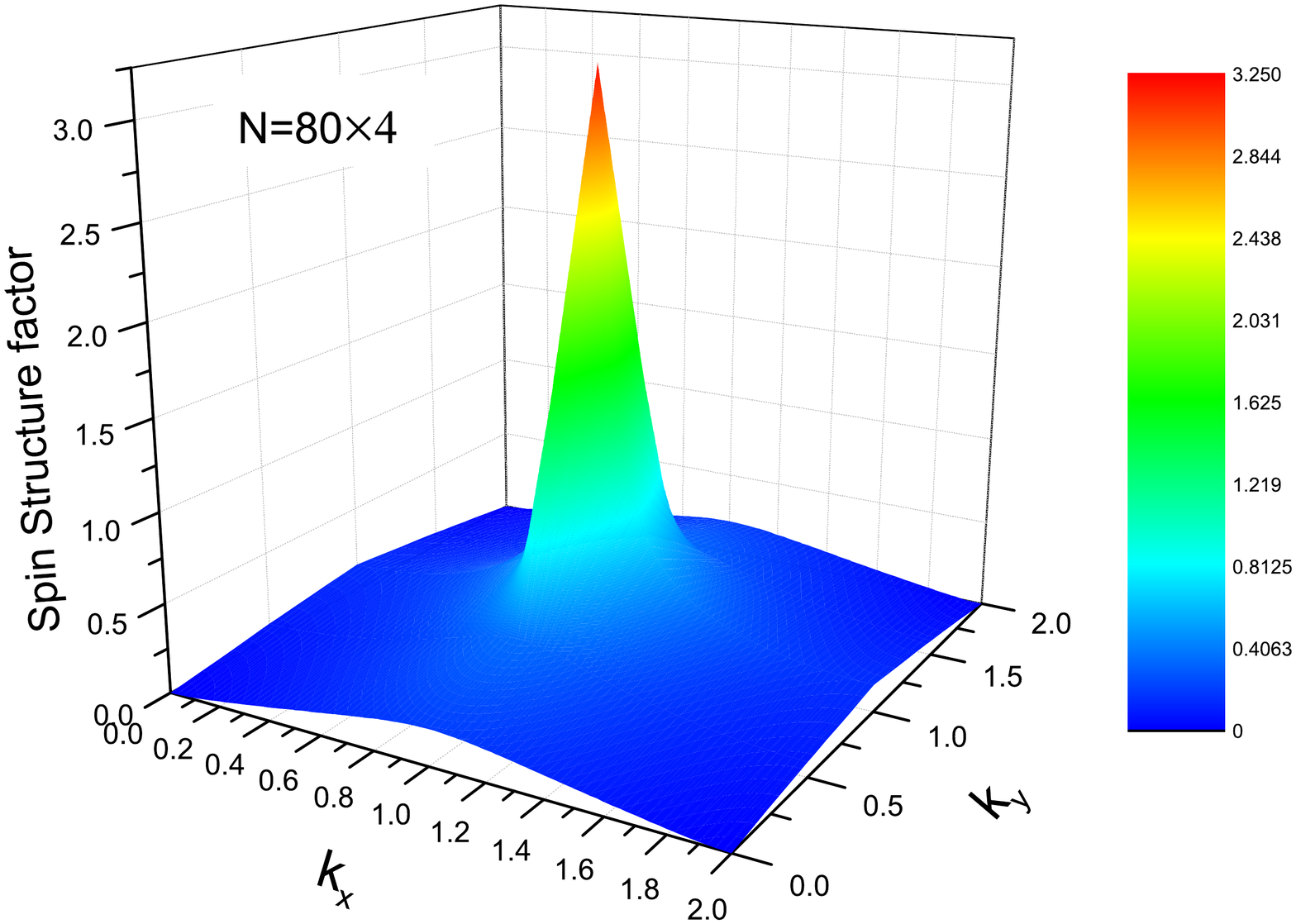}
\end{center}
\renewcommand{\figurename}{Fig.}
\caption{Spin structure factor for the 4-leg ladder
($N$=80$\times4$). The peak is located at ($\pi$,$\pi$),
indicating strong spin antiferromagnetic correlations even though a
finite spin gap is present. } \label{structurefactor}
\end{figure}

{\it Some details of the hole density distribution.} --- { Fig.}~\ref{Total_Contour} presents
the contour plot of $\langle {n_i^h } \rangle$ for different ladders, $N=40\times N_y$ with $N_y=2$, $3$,
$4$, and $5$. For each contour plot, we have checked
numerically that the sum rule is satisfied, i.e.,
$\sum_i\langle {n_i^h } \rangle=1$, where the summation is over all the sites.

\begin{figure}
\begin{center}
\includegraphics[width=0.45\textwidth]{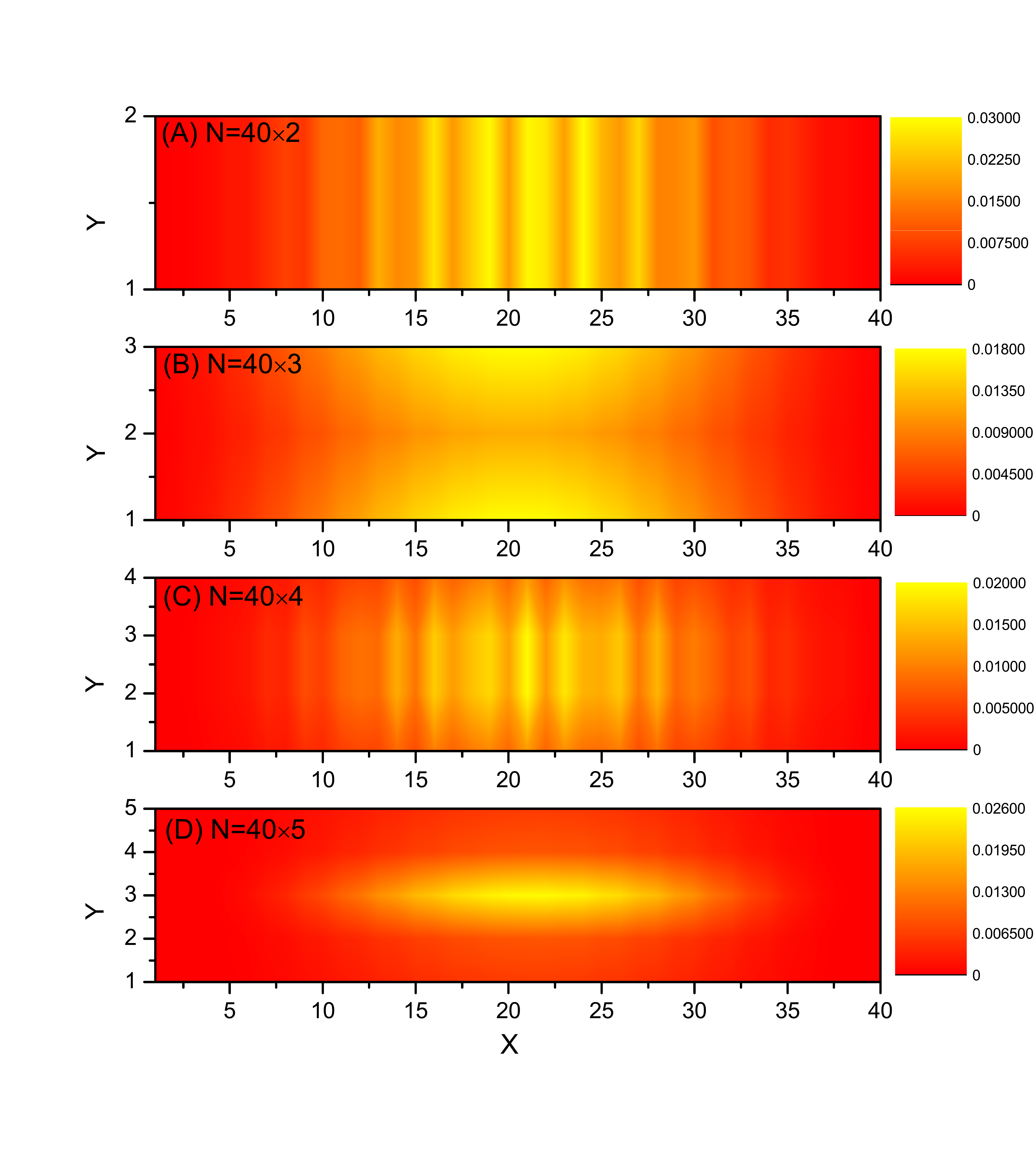}
\end{center}
\renewcommand{\figurename}{Fig.}
\caption{Contour plots of the hole distribution in real space. The
sample size $40\times2$ (A), $40\times3$ (B), $40\times4$ (C),
and $40\times5$ (D) from top down. } \label{Total_Contour}
\end{figure}

A prominent feature can be clearly seen in these contour plots
(see also { Fig.}~\ref{ratio-hole} (B)).
That is, for even-leg ladders there is a spatial oscillation on top
of the localized density profile, while it is absent
for odd-leg ladders. We wish to point out that this is
concomitant with the parity effect of the spin-spin correlation
shown in { Fig.}~\ref{spincorrelation}. In fact,
the spin-gap effect is reduced as the hopping integral $t$ increases,
and eventually the parity effect disappears in the large $t/J$ limit. Accordingly, the
spatial oscillation of $\left \langle {n_{i}^{h}}\right \rangle $ must
be diminished also in this limit. This prediction has been fully confirmed by
numerical simulations, see { Fig.}~\ref{ratio-hole} (B). In addition,
we have observed that the localization length
{ monotonically} decreases as the ratio $t/J$ increases, see the insets of
{ Fig.}~\ref{ratio-hole}. This suggests that the spin dynamics, governed by
the superexchange $J$, { does not play important} roles in establishing {
the observed} self-localization.

\begin{figure}[tbp]
\begin{center}
\includegraphics[width=0.45\textwidth]{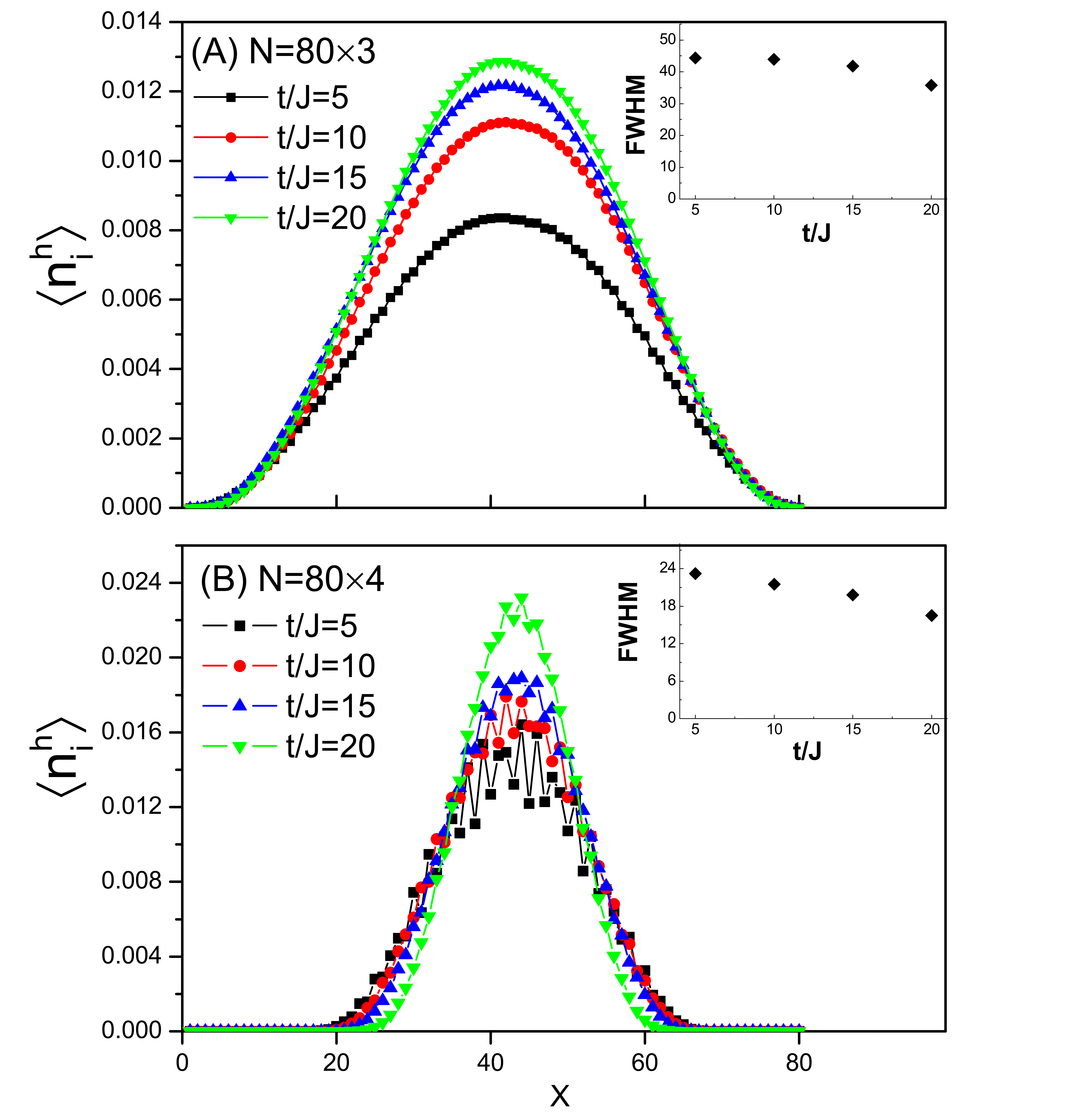}
\end{center}
\par
\renewcommand{\figurename}{Fig.}
\caption{
The hole density profile
$\langle n_{i}^{h}\rangle$ (with $i$ located on the middle chain)
at different ratios of $t/J$. The ladder size
is $80\times 3$ (A) and $80\times 4$ (B). The localization length
(the FWHM) is found to decrease as $t/J$ increases (insets).
Moreover, the amplitude of the spatial oscillation in the $4$-leg ladder is suppressed
as $t/J$ increases, indicating that the behavior of
even- and odd-leg ladders converges in the large $t/J$ limit.}
\label{ratio-hole}
\end{figure}

{\it Electron and hole momentum distributions.} --- { Fig.}~\ref{n(k)contour}
shows the contour plot of the hole momentum distribution function
$1-n(\textbf{k})$ for the $3$-leg (A) and $4$-leg (B) ladders, respectively at $N_x=80$. The maximum of
$1-n(\textbf{k})$ appears at $k_y$=$\pm \pi/2$ for the $4$-leg ladder and
$k_y$=$\pm2\pi/3$ for the $3$-leg ladder. The $t/J$-dependence of
the hole momentum distribution is shown in
{ Fig.}~\ref{ratio}. They show that the momentum distribution jump is { monotonically} reduced
with the increase of $t/J$ .

\begin{figure}
\renewcommand{\thesubfigure}{(\Alph{subfigure})}
\centering
\subfigure[N=80$\times$3]{
\includegraphics[width=0.45\textwidth]{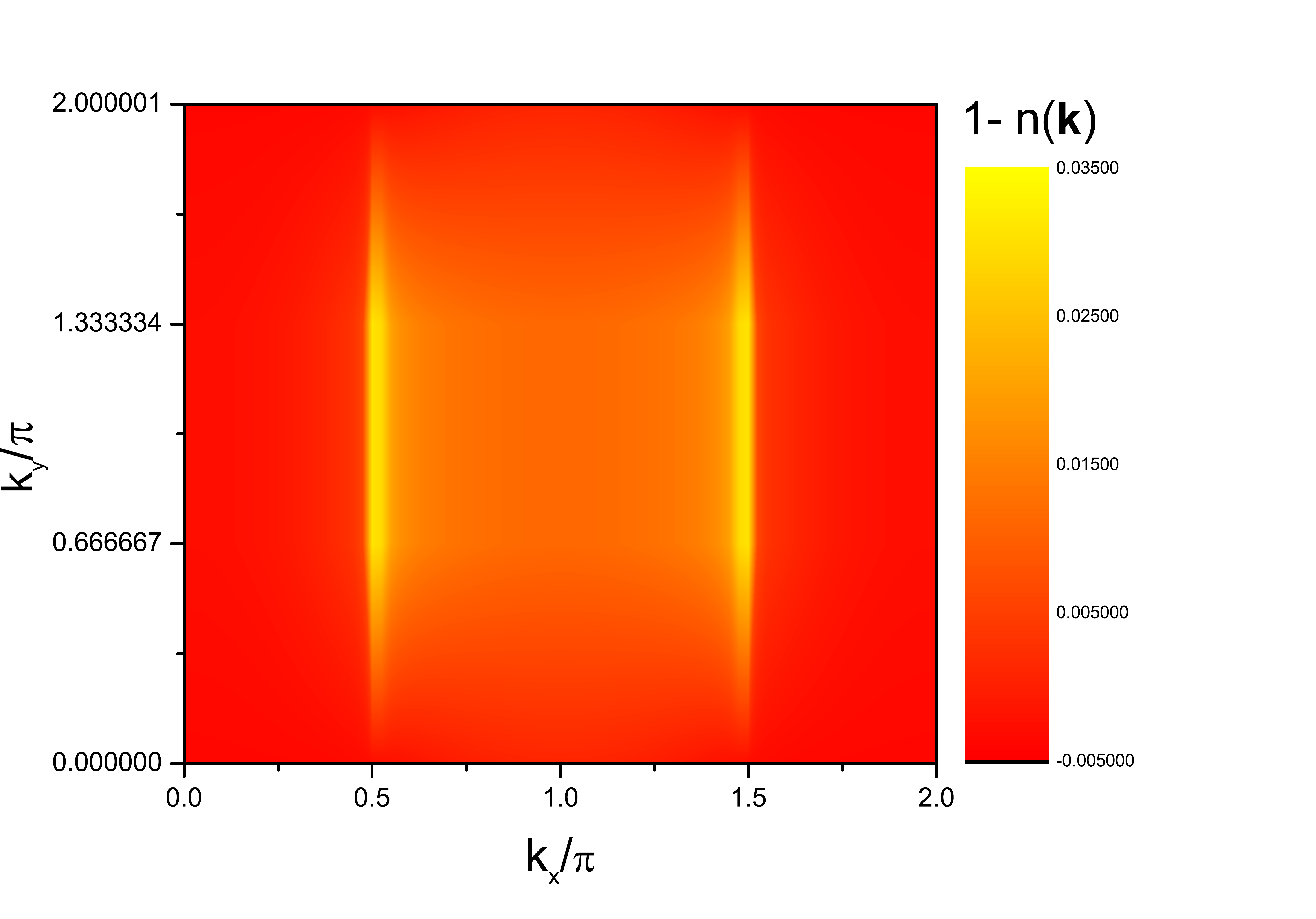}
}
\subfigure[N=80$\times$4]{
\includegraphics[width=0.45\textwidth]{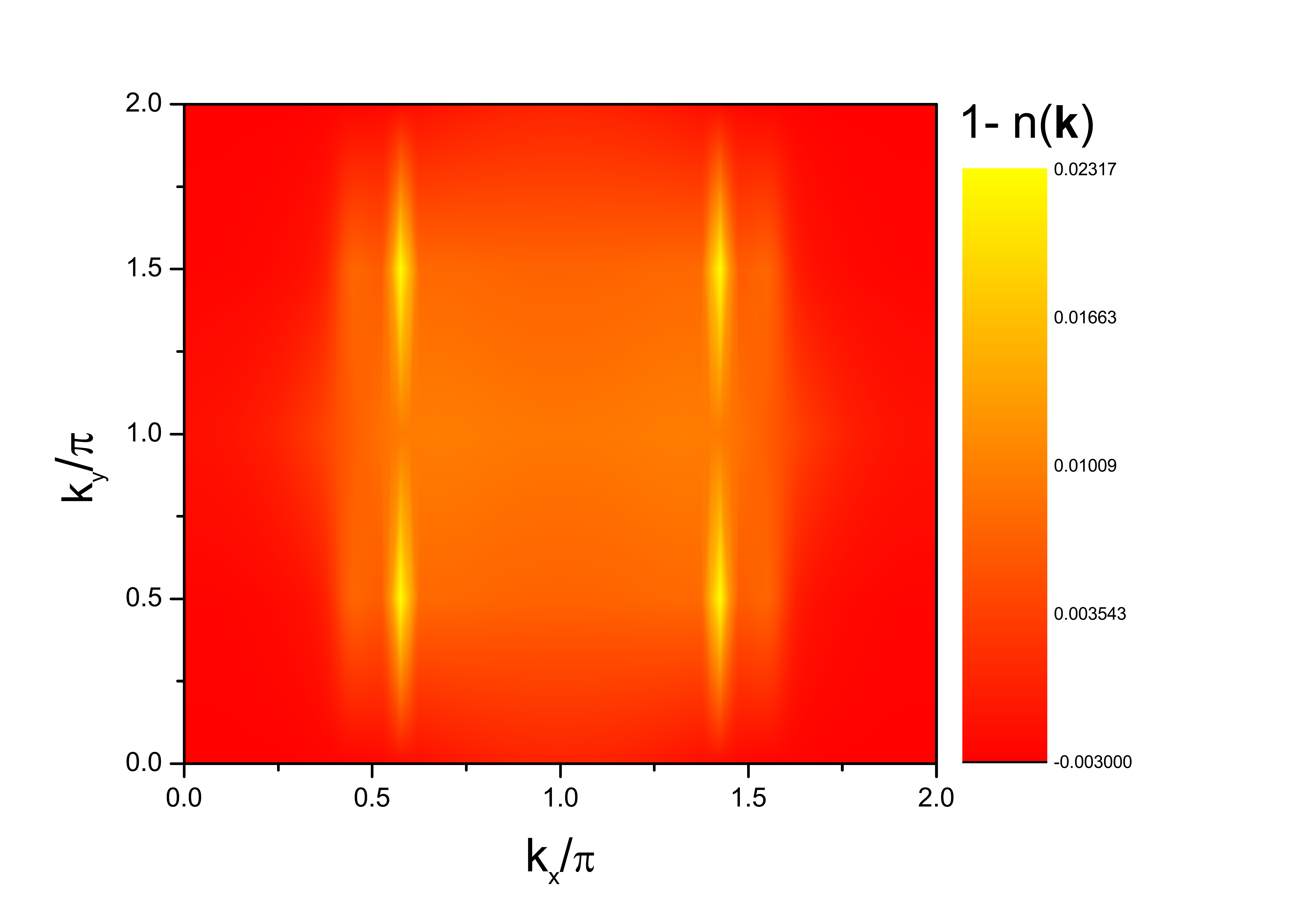}
}
\renewcommand{\figurename}{Fig.}
\caption{Contour plots of the hole momentum distribution $1-n({\bf k})$
for $3$-leg (A) and $4$-leg (B) ladders. } \label{n(k)contour}
\end{figure}

\begin{figure}[tbp]
\begin{center}
\includegraphics[width=0.9\textwidth]{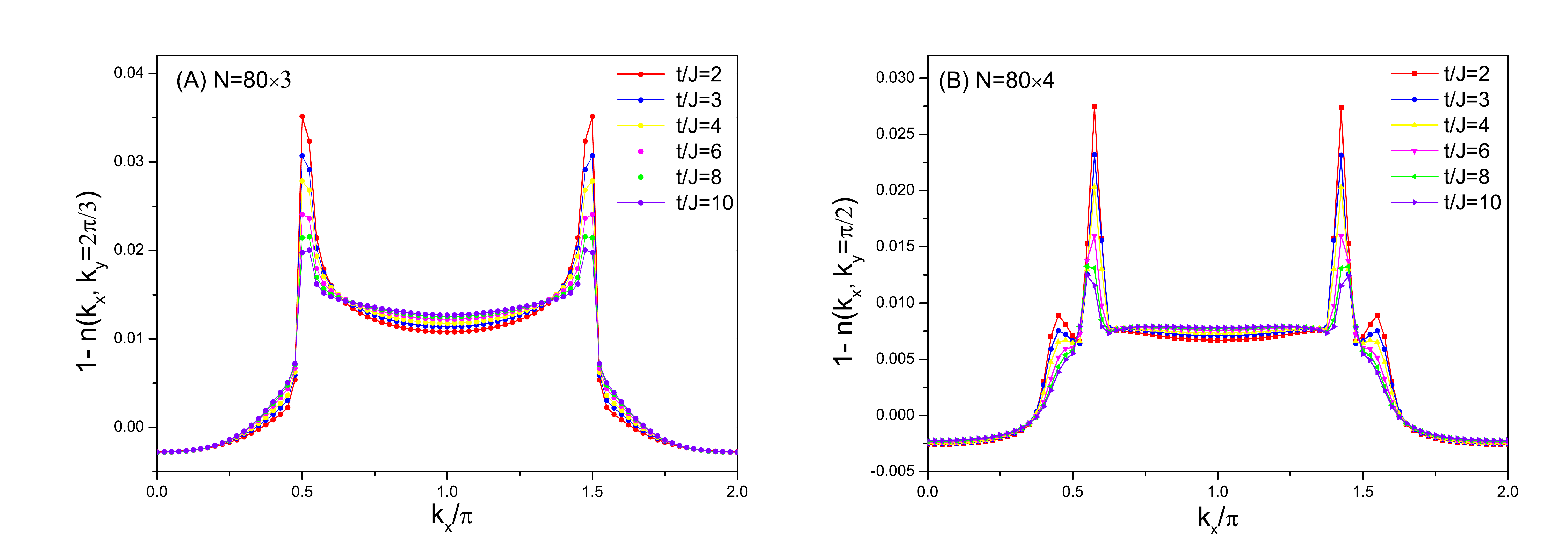}
\end{center}
\par
\renewcommand{\figurename}{Fig.}
\caption{The hole momentum distribution at different ratios of
$t/J$ for $3$-leg (A) and $4$-leg (B) ladders.  }
\label{ratio}
\end{figure}

\end{widetext}

\end{document}